\documentstyle[preprint,aps]{revtex}

\begin{document}

\tighten

\draft

\title{The Holographic Supercurrent Anomaly}

\author{M. Chaichian\renewcommand{\thefootnote}{\dagger}\thanks{
E-mail: Masud.Chaichian@helsinki.fi} and W.F.
Chen\renewcommand{\thefootnote}{\ddagger}\thanks{E-mail:
Wen-feng.Chen@helsinki.fi} }

\address{ High Energy Physics Division, Department of Physical Sciences,
University of Helsinki\\
and\\
 Helsinki Institute of Physics, FIN-00014,  Helsinki,
Finland}

\maketitle

\begin{abstract}
The $\gamma$-trace anomaly of supersymmetry current in a
supersymmetric gauge theory  shares a superconformal anomaly
multiplet with the  chiral $R$-symmetry anomaly and the Weyl
anomaly, and its holographic reproduction is a valuable test to
the AdS/CFT correspondence conjecture. We investigate how the
$\gamma$-trace anomaly of the supersymmetry current of ${\cal
N}=1$ four-dimensional supersymmetric gauge theory in an ${\cal
N}=1$ conformal supergravity background can be extracted out from
the ${\cal N}=2$ gauged supergravity in five dimensions. It is
shown that the reproduction of this super-Weyl anomaly
 originates from the following two facts: First the ${\cal N}=2$
bulk supersymmetry transformation converts into ${\cal N}=1$
superconformal transformation on the boundary, which consists of
${\cal N}=1$ supersymmetry transformation and special conformal
supersymmetry (or super-Weyl) transformation; second the
supersymmetry variation of the bulk action
 of five-dimensional gauged supergravity is a total derivative.
  The non-compatibility
 of supersymmetry and  the super-Weyl transformation invariance yields
 the holographic supersymmetry current anomaly. Furthermore,
 we speculate on that the contribution from the external gauge
 and gravitational background fields to the
 superconformal anomaly may have different holographic origin.

\vspace{3mm}

\end{abstract}

\vspace{3ex}

\section{Introduction}

  The AdS/CFT correspondence conjecture \cite{mald}
  states that  the type IIB string theory compactified on $AdS_5\times S^5$
  theory with $N$ units of $R-R$ flux on $S^5$ describes the same physics
  as ${\cal N}=4$ $SU(N)$ supersymmetric Yang-Mills theory.
  Further, the explicit definition was  given
  in Refs.\,\cite{gkp} and \cite{witt1} as the following. Given
 the type IIB superstring theory in the background
$AdS_{d+1}\times X^{9-d}$, with $X^{9-d}$ being a compact Einstein
manifold, the boundary effect of the type IIB superstring  theory
must be considered since $AdS_{d+1}$ has a boundary at spatial
infinity, which is actually a copy of Minkowski space $M_d$ (or
its compactified version $S_d$).  The partition function of such a
string theory with certain boundary value of a bulk field takes
the following form,
\begin{eqnarray}
\left.Z_{\rm
String}\left[\phi\right]\right|_{\phi\rightarrow\phi_{0}}
=\int_{\phi (x,0)=\phi_{0}(x)}{\cal D}\phi (x,r) \exp\left(-S[\phi
(x,r)]\right), \label{spf}
\end{eqnarray}
where $\phi^{(0)}(x)$ is the boundary value of the $AdS_{d+1}$
bulk quantity $\phi (x,r)$ such as the graviton, gravitino,
$NS-NS$ and $R-R$ antisymmetric tensor fields etc.. Since the
isometry group $SO(2,d)$ of $AdS_{d+1}$ acts as the conformal
group  on $M_d$, and hence a quantum field theory defined on the
boundary should be conformal invariant. The $AdS/CFT$
correspondence conjecture means that the type IIB superstring
partition function (\ref{spf}) should be identical to the
generating functional for the correlation functions of the
composite operators of certain conformal field theory
\begin{eqnarray}
Z_{\rm CFT}\left[\phi_{0}\right]&=& \left\langle \exp\int_{M^d}
d^dx
{\cal O} (x)  \phi_{0}(x)\right\rangle \nonumber\\
&=&\sum_n\frac{1}{n!}\int \prod_{i=1}^n d^d x_i \left\langle {\cal
O}_1 (x_1)\cdots {\cal O}_n (x_n) \right\rangle \phi_{0} (x_1)
\cdots \phi_{0} (x_n)\nonumber\\
&{\equiv}& \exp\left(-\Gamma_{\rm CFT} [\phi_{0}]\right),
\label{acc1}
\end{eqnarray}
$\Gamma [\phi_{0}]$ being the quantum effective action describing
the composite operators  interacting with $\phi_{0}$ background
field. That is,
 \begin{eqnarray}
   \left.Z_{\rm
String}\left[\phi\right]\right|_{\phi\rightarrow\phi_{0}}
   =Z_{\rm CFT}\left[\phi_{0}\right] \,.
   \label{acc2}
 \end{eqnarray}
In the large-$N$ case,  the type IIB string correction to
supergravity is proportional to  $1/\sqrt{g_sN}$, $g_s$ being the
string coupling, thus one can neglect the string effect and just
consider its low-energy effective theory, the type IIB
supergravity. In this case, the  partition function of the type
IIB superstring can be evaluated as the exponential of the
supergravity action in a field configuration $\phi^{\rm
cl}[\phi^{0}]$ which satisfies the classical equation of motion of
the supergravity with the boundary condition given by $\phi_{0}$,
i.e.,
\begin{eqnarray}
\left.Z_{\rm
String}\left[\phi\right]\right|_{\phi\rightarrow\phi_{0}}
 =\exp\left(-S_{\rm SUGRA}[\phi^{\rm cl}[\phi_{0}]]\right) \,.
\label{acc3}
\end{eqnarray}
Comparing \,(\ref{acc3}) with (\ref{acc1}) and (\ref{acc2}), we
immediately conclude that the background effective action of the
large-$N$ limit of the $d$-dimensional conformal field theory can
be approximately equal to the on-shell classical action of
$AdS_{d+1}$ supergravity with non-empty boundary,
\begin{eqnarray}
\Gamma_{\rm CFT}[\phi_{0}]=S_{\rm SUGRA}[\phi^{\rm
cl}[\phi_{0}]]=\int d^d x \sum_{n}\phi_0^{(n)} (x) \left\langle
{\cal O}^{(n)}\right\rangle. \label{acc4}
\end{eqnarray}
In the above equation, ${\cal O}^{(r)}$ are the various composite
operators in the superconformal field theory such as the
energy-momentum tensor and chiral $R$-symmetry current etc. and
$\phi_0^{(r)}$ are the corresponding background fields such as the
gravitational and gauge  fields etc., which are boundary values of
the corresponding bulk fields.

Let us emphasize the role of the five-dimensional gauged
supergravities \cite{gst,awada,guna2} in AdS/CFT correspondence
\cite{ferr}.  The $AdS_5\times S^5$ background arises from the
near horizon limit of $D3$-brane solution of type IIB supergravity
\cite{agmo}. In the $AdS_5\times S^5$ background, the spontaneous
compactification on $S^5$ of the type IIB  supergravity actually
occurs \cite{freu,duff}. With the assumption that there exists a
consistent nonlinear truncation of the massless modes from the
whole Kluza-Klein spectrum of the type IIB supergravity
compactified on $S^5$ \cite{duff,marcus,kim}, the resultant theory
should be the $SO(6)(\cong SU(4))$ gauged ${\cal N}=8$ $AdS_5$
supergravity since the isometry group $SO(6)$ of the internal
space $S^5$ becomes the gauge group of the compactified theory and
the $AdS_5\times S^5$ background preserves all of the
supersymmetries of type IIB supergravity \cite{guna2}.
Furthermore, if the background for the type IIB supergravity is
$AdS_5 \times X^5$ with $X^5$ being an Einstein manifold rather
than $S^5$ such as $T^{1,1}=(SU(2)\times SU(2))/U(1)$ or certain
orbifold, then due to the singularities in the internal manifold,
the number of preserved supersymmetries in the compactified
$AdS_5$ supergravity is reduced and the isometry group of the
theory also changes \cite{roman2,kw2,kach,law}.  One can thus
obtain the gauged ${\cal N}=2,4$ $AdS_5$ supergravities in five
dimensions, and their dual field theories  are believed to be
$N=1,2$ supersymmetric gauge theories \cite{ferr,kw2,kach,law}. In
a strict sense, a supersymmetric gauge theory with lower
supersymmetries is not a conformal invariant theory since its beta
function does not vanish. However, it was shown that
renormalization group flow of this type of supersymmetric gauge
theory has the fixed point, at which the conformal invariance can
arise \cite{kach,law,shif,seib,lei2,early1,early2}. The $AdS/CFT$
correspondence between the ${\cal N}=2,4$ gauged supergravities in
five dimensions and  ${\cal N}=1,2$ supersymmertric gauge theories
can thus be established \cite{ferr,kw2,kach,law}.

With the truncation  from the Kaluza-Klein tower,
Eq.\,(\ref{acc4}) can be
 considered as a quantum effective action
describing a superconformal gauge theory in an external
supergravity background \cite{witt1,liu}, only where the external
fields are provided by the boundary values of those in a
one-dimension-higher bulk space-time. It is well known that the
superconformal  anomaly will arise from a classical superconformal
gauge theory  in an external supergravity background. The
reproduction of the  superconformal anomaly from the bulk gauged
supergravity will provide an important support to the above
$AdS/CFT$ correspondence conjecture  at low-energy level.

In the next section, we shall briefly introduce the superconformal
anomaly for the ${\cal N}=1$ supersymmetric gauge theory in an
external ${\cal N}=1$ conformal  supergravity background. Section
3 is devoted to a review of the ${\cal N}=2$ gauged supergravity
in five dimensions and its $AdS_5$ classical solution.
 We also emphasize  the behaviour of the fields
of gauged supergravity near the $AdS_5$ boundary and the reduction
of bulk supersymmetry transformation into an ${\cal N}=1$
superconformal transformation on the $AdS_5$ boundary. It is well
known that the ${\cal N}=1$ superconformal transformation consists
of the ${\cal N}=1$ four-dimensional supersymmetry transformation
and the special conformal supersymmetry transformation (or
super-Weyl in curved space-time background). In section 4, we
calculate the supersymmetry variation of the five-dimensional
${\cal N}=2$ gauged supergravity and extract out the surface term.
Furthermore, considering the fact that the four-dimensional
supersymmetry and the super-Weyl symmetry cannot be preserved
simultaneously, we find the  external gauge field part of the
holographic super-Weyl anomaly of ${\cal N}=1$ supersymmetric
gauge theory  from the boundary term the five-dimensional gauged
supergravity. In Section 5 we  summarize the main results and
present some further discussions. Our results suggest that the
contributions to the superconformal anomaly from the external
vector field and gravitational field may have different
holographic origin.
%the external vector contribution lies in the
%leading term in the near-boundary expansion of the bulk fields, while
%the gravitational background contribution comes from the terms
%with logarithmic dependence on the radial coordinate in the expansion.

\section{Superconformal Anomaly Multiplet in External
Conformal Supergravity Background}

A general four-dimensional ${\cal N}=1$ supersymmetric $SU(N)$
gauge theory consists of  the ${\cal N}=1$ supersymmetric
Yang-Mills theory coupled with ${\cal N}=1$ massless matter fields
in various possible representations of the gauge group. The
classical Lagrangian density is
\begin{eqnarray}
{\cal L}_{\rm }= \mbox{Tr}\left(-\frac{1}{4}W_{\mu\nu} W^{\mu\nu}
+\frac{1}{2}i\overline{\lambda}\gamma^\mu \nabla_\mu
\lambda\right) +{\cal L}_{\rm matter},
\end{eqnarray}
where $W_{\mu\nu}$ is the field strength for the $SU(N)$ gauge
field and the $\lambda$ is a Majorana spinor in the adjoint
representation of $SU(N)$. Due to the supersymmetry, its
energy-momentum tensor $\theta^{\mu\nu}$, the supersymmetry
current $s^{\mu}$ and the axial vector (or equivalently chiral)
R-current $j^{(5)\mu}$
 lie in a supermultiplet \cite{ilio,anm}.
These currents at classical level are not only conserved,
\begin{eqnarray}
\partial_\mu \theta^{\mu\nu}=\partial_\mu s^\mu=\partial_\mu
j^{(5)\mu}=0,
\end{eqnarray}
but also satisfy further algebraic constraints
\begin{eqnarray}
\theta^{\mu}_{~\mu}=\gamma_\mu s^\mu=0. \label{superconf}
\end{eqnarray}
This will promote  the Poincar\'{e} supersymmetry to a
superconformal symmetry since one can construct three more
conserved currents,
\begin{eqnarray}
d^\mu {\equiv} x_\nu \theta^{\nu\mu}, ~ k_{\mu\nu}{\equiv} 2 x_\nu
x^\rho\theta_{\rho\mu}-x^2\theta_{\mu\nu}, ~ l_\mu{\equiv} ix^\nu
\gamma_\nu s_\mu.
\end{eqnarray}
These three new conserved currents give the generators for
dilatation, conformal boost and special supersymmetry
transformation.
 However,  the superconformal symmetry may become
 anomalous at quantum level. In the case that all of them, the trace of
 energy-momentum tensor, $\theta^\mu_{~\mu}$, the $\gamma$-trace
 of supersymmetry current, $\gamma^\mu s_\mu$ and the divergence
 of the chiral $R$-current, $\partial_\mu j^{(5)\mu}$,  get
 contribution from quantum effects,
\begin{eqnarray}
\left(\partial_\mu j^{(5)\mu},\gamma^\mu s_\mu,
\theta^{\mu}_{~\mu}\right) \label{cas}
\end{eqnarray}
 will form a (on-shell) chiral supermultiplet with the $\partial_\mu
j^{(5)\mu}$ playing the role of the lowest component of the
corresponding composite chiral superfield
\cite{anm,sibold,grisaru}.

In general, there are two possible sources for above
 chiral supermultiplet anomaly \cite{grisaru}. One is due to the
non-vanishing beta function of ${\cal N}=1$ supersymmetric
Yang-Mills theory; The other one comes from
  the coupling of  above supercurrent multiplet with
the external supergravity fields. In this paper, we shall
concentrate on the superconformal anomaly arising from the latter
one. In this case, the $\gamma$-trace anomaly consists only of the
super-Weyl anomaly since the corresponding special conformal
supersymmetry transformation is just the supersymmetric analog of
the Weyl (or local scale) transformation.
  Note that in a supersymmetric gauge theory,
the Poincar\'{e}
 symmetry corresponding to the energy-momentum tensor $\theta_{\mu\nu}$,
 the supersymmetry corresponding to the supersymmetry current
 $s_\mu$, and the chiral $R$-symmetry to the axial vector
 current $j_\mu^{(5)}$ are all
 global symmetries and there no
 gauge fields within the supersymmetric gauge theory itself to
 couple with them. If there are
 some external supergravity fields $g_{\mu\nu}$, axial vector fields $A_\mu$
 and vector-spinor (Rarita-Schwinger) fields $\psi_{\mu}$ couple
 to $\theta^{\mu\nu}$,  $j_\mu^{(5)}$ and $s_{\mu}$,
 respectively,
 \begin{eqnarray}
{\cal L}_{\rm ext}=\int d^4x \sqrt{-g}\left(
g_{\mu\nu}\theta^{\mu\nu} + A_\mu j^{(5)\mu}+\overline{\psi}_\mu
s^{\mu}\right), \label{clag}
\end{eqnarray}
 there will arise  external superconformal
anomaly chiral supermultiplet. The action (\ref{clag}) describing
the coupling of the external supergravity fields with the currents
of supersymmetric Yang-Mills theory shows  that the covariant
conservations  of the currents, $\nabla_\mu
\theta^{\mu\nu}=\nabla_\mu s^\mu=0$, are equivalent to the local
gauge transformation invariance of the external supergravity
system,
\begin{eqnarray}
\delta g_{\mu\nu} (x) &=& \nabla_\mu \xi_\nu +\nabla_\nu
\xi_\mu ,\nonumber\\
\delta\psi_\mu (x)&=& \nabla_\mu \chi (x).
 \label{egt1}
\end{eqnarray}
Furthermore, the covariant conservation of the axial vector
current $j_\mu^{(5)}$ and
 the vanishing of both the $\gamma$-trace of supersymmetry current
and the trace of energy-momentum  tensor at classical level,
\begin{eqnarray}
\nabla_\mu j^{(5)\mu}=\gamma^\mu s_\mu =\theta^{\mu}_{~\mu}=0,
\label{supercon}
\end{eqnarray}
mean the Weyl transformation invariance of $g_{\mu\nu}$, the
super-Weyl symmetry and the $U(1)$ chiral gauge symmetry of the
corresponding external supergravity system,
\begin{eqnarray}
\delta g_{\mu\nu}&=&g_{\mu\nu} \sigma (x),\nonumber\\
\delta \psi_\mu &=& \gamma_\mu \eta (x),\nonumber\\
\delta A_\mu (x) &=& \partial_\mu \Lambda (x). \label{busy}
\end{eqnarray}
The transformations (\ref{egt1}) and (\ref{busy}) imply that the
external fields
\begin{eqnarray}
(g_{\mu\nu}, \psi_\mu, A_\mu)
\end{eqnarray}
 constitute
an off-shell ${\cal N}=1$ conformal supergravity multiplet
\cite{kaku,fra}. Therefore, in the context of the $AdS/CFT$ (or
more generally gravity/gauge) correspondence the
 superconformal anomaly   in ${\cal N}=1$
supersymmetric gauge theory due to the supergravity external
sources will be reflected in the explicit violations of the bulk
symmetries of ${\cal N}=2$ gauged $AdS_5$ supergravity on the
boundary \cite{witt1,hesk,bian1}.

With no consideration on the quantum correction from the dynamics
of the supersymmetric gauge theory, the external superconformal
anomaly is exhausted  at one-loop level. The external
superconformal anomaly multiplet for  ${\cal N}=1$ pure
supersymmetric Yang-Mills theory is listed as the following
\cite{grisaru}:
\begin{eqnarray}
\nabla_\mu j^{(5)\mu} &=&\frac{N^2-1}{16\pi^2}
\left(-\frac{1}{24}{R}_{\mu\nu\lambda\rho}\widetilde{R}^{\mu\nu\lambda\rho}
+ F_{\mu\nu}\widetilde{F}^{\mu\nu}\right),
\nonumber\\
 \gamma_\mu s^\mu &=&\frac{N^2-1}{16\pi^2}\left(\frac{1}{16}
 R^{\mu\nu\lambda\rho}\gamma_{\lambda\rho}
 +\frac{1}{8}F_{\mu\nu}\right)\left(D_\mu \psi_\nu-
 D_\nu \psi_\mu\right),
\nonumber\\
\theta^\mu_{~\mu} &=&
\frac{N^2-1}{16\pi^2}\left(\frac{1}{8}C_{\mu\nu\lambda\rho}
C^{\mu\nu\lambda\rho}-\frac{3}{16}
\widetilde{R}_{\mu\nu\lambda\rho}\widetilde{R}^{\mu\nu\lambda\rho}
+\frac{1}{3}F_{\mu\nu} F^{\mu\nu}\right). \label{exone}
\end{eqnarray}
%where $G$ is the square of the four dimensional Weyl tensor and
%$G$ is the Euler number density,
%\begin{eqnarray}
%F_{\mu\nu} &=&\partial_\mu A_\nu-\partial_\nu A_\mu, \nonumber\\
%H&=& C_{\mu\nu\lambda\rho}C^{\mu\nu\lambda\rho}=
%R_{\mu\nu\lambda\rho}R^{\mu\nu\lambda\rho} -2 R_{\mu\nu}
%R^{\mu\nu}+\frac{1}{3}R^2, \nonumber\\
%G &=&
%\widetilde{R}_{\mu\nu\lambda\rho}\widetilde{R}_{\mu\nu\lambda\rho}
%=R_{\mu\nu\lambda\rho}R_{\mu\nu\lambda\rho}-4R_{\mu\nu}
%R^{\mu\nu}+R^2
%\end{eqnarray}
In above equations,
$\gamma_{\mu\nu}=i/2\,[\gamma_\mu,\gamma_\nu]$;
$F_{\mu\nu}=\partial_\mu A_\nu- \partial_\nu A_\mu$  is the field
strength corresponding to the external $U_R(1)$ vector field
$A_\mu$; $R_{\mu\nu\lambda\rho}$ and $C_{\mu\nu\lambda\rho}$  are
the Riemannian and Weyl tensors corresponding to the gravitational
background field $g_{\mu\nu}$; $D_\mu$ is the covariant derivative
with respect to both the external gravitational and gauge fields.
The factor $N^2-1$ comes from the fact that the gauginoes are in
the adjoint representation
 of $SU(N)$ gauge group and hence there are $N^2-1$ copies. Note that
Eq.\,(\ref{exone}) contains only the contribution to the
superconformal anomaly coming from the vector multiplet of ${\cal
N}=1$ supersymmetric $SU(N)$ Yang-Mills theory. Naturally, there
will also arise the contribution from the ${\cal N}=1$
 chiral supermutiplets if they are present in the theory. Their contributions
 to the anomaly coefficient will be proportional to $N$ or $N^2-1$.
 This depends on the chiral multiplets belonging to the fundamental
or adjoint representations of $SU(N)$ group.

In the context of the AdS/CFT correspondence, the holographical
arising of the chiral $U_R(1)$ anomaly was pointed out by Witten
that it should come directly from the Chern-Simons term in the
five-dimensional gauged supergravity \cite{witt1}. Further, the
holographic Weyl anomaly was shown in Ref.\,\cite{hesk} through a
procedure called holographic renormalization and in
Ref.\,\cite{imbi} by the holomorphic dimensional regularization
and a special bulk diffeomorphism preserving the Fefferman-Graham
metric \cite{feff} of an arbitrary $d+1$-dimensional manifold with
boundary topologically isomorphic to $S^d$, respectively.
Actually, there have arisen quite a number of papers to discuss
various aspects of the holographic Weyl anomaly including the
modified models \cite{noji}, the asymptotically $AdS$ space-time
\cite{kraus}, the subleading order of $1/N$ correction
\cite{ahar,blau}
 and new calculation
framework such as the Hamilton-Jacobi equation \cite{fuku} etc.
The aim of this paper is to tackle the holographic origin of the
supersymmetry current anomaly $\gamma_\mu s^\mu$.

\section{ ${\cal N}=2$ Gauged Supergravity   in Five Dimensions
and Its $AdS_5$ Boundary Reduction}

To show clearly how the holographic supersymmetry current anomaly
arises, we shall review some typical features of the ${\cal N}=2$
gauged supergravity in five dimensions.

The ungauged ${\cal N}=2$ supergravity in five dimensions has the
same structure as ${\cal N}=1$ eleven-dimensional supergravity
\cite{crem1}. It has a global $USp(2)\cong SU(2)$ R-symmetry, and
it contains a graviton $E_M^{~A}$, two gravitini $\Psi_M^i$ and a
vector field ${\cal A}_M$ \cite{gst,crem2,guna}, $M, A=0,\cdots,5$
are the Riemannian and local Lorentz indices, respectively, and
$i=1,2$ the $SU(2)$ doublet indices. The gravitini are the
$USp(2)\cong SU(2)$ doublets and the symplectic Majorana spinors.
The classical Lagrangian density takes a simple form \cite{crem2},

\begin{eqnarray}
{E}^{-1} \widetilde{\cal L} &=& -\frac{1}{2}{\cal R}[\Omega(E)]
-\frac{1}{2}\overline{\Psi}_M^i {\Gamma}^{MNP}{\nabla}_N {\Psi}_{P
i} -\frac{1}{4} a_{00}{\cal F}_{MN}{\cal F}^{MN}
\nonumber\\
&& -\frac{3}{8}\sqrt{\frac{1}{6}}ih_0 \left(
\overline{\Psi}_M^i\Gamma^{MNPQ} \Psi_{N i}{\cal
F}_{PQ}+2\overline{\Psi}^{M i}\Psi^N_i F_{MN}\right)
\nonumber\\
&& +\frac{C}{6\sqrt{6}}{E}^{-1}\epsilon^{MNPQR} {\cal F}_{MN}{\cal
F}_{PQ} {\cal A}_R +\mbox{four-fermi terms},
 \end{eqnarray}
where the covariant derivative on the spinor field is defined with
the spinor connection $\Omega_{M}^{~AB}$,
\begin{eqnarray}
\nabla_M \Psi_N^i=\left(\partial_M
+\frac{1}{4}\Omega_{M}^{~AB}\Gamma_{AB} \right)\Psi_N^i.
 \end{eqnarray}

 The gauging of above supergravity is just turning the $U(1)$ subgroup of
 the global
 $SU(2)$ R-symmetry group into a local gauge group and straightforwardly
 considering the vector
 field as the $U(1)$  gauge field \cite{gst}. The space-time
 covariant derivative  on the gravitini
 will be enlarged to include the $U(1)$ gauge covariant derivative,
\begin{eqnarray}
D_M\Psi_N^i=\nabla_M \Psi_N^i+g {\cal A}_{M}\delta^{ij}\Psi_{N j}.
\end{eqnarray}
The gauged ${\cal N}=2$ supergravity action is
\begin{eqnarray}
E^{-1} {\cal L} ={E}^{-1} \widetilde{\cal L} +g^2
P_0^2-\frac{i\sqrt{6}}{8}g
\overline{\Psi}_M^i\Gamma^{MN}\Psi^{j}_N \delta_{ij}P_0.
\label{gaugedf}
\end{eqnarray}
The gauged ${\cal N}=2$ supergravity  has  $AdS_5$ classical
solution that preserves ${\cal N}=2$ supersymmetry with the
cosmological constant proportional to $P_0$ \cite{gst}. To make
the $AdS_5$ classical solution take the standard form,
\begin{eqnarray}
   ds^2 &=& \frac{l^2}{r^2}\left[g_{\mu\nu}(x,r)dx^\mu dx^\nu
   -\left(dr\right)^2\right],
 \nonumber\\
   {\cal A}_M &=&\Psi_M=0,
   \label{metrican}
   \end{eqnarray}
and the full ${\cal N}=2$ supersymmetry preserved in this
background, one must choose the parameters in the Lagrangian
(\ref{gaugedf}) as the following ones \cite{gst,bala}:
 \begin{eqnarray}
g=\frac{3}{4},
~~h_0=\frac{l}{2}\sqrt{\frac{3}{2}},~~h^0=\frac{1}{h_0}, ~~V_0=1,
~~P_0=2 h^0 V_0=\frac{4}{l}\sqrt{\frac{2}{3}}, ~~
a_{00}=\left(h_0\right)^2=\frac{3l^2}{8}.
\end{eqnarray}
Consequently, the  Lagrangian density (\ref{gaugedf})
  up to the quadratic terms in spinor fields becomes
 \begin{eqnarray}
E^{-1} {\cal L} &=& -\frac{1}{2}{\cal
R}-\frac{1}{2}\overline{\Psi}_M^i \Gamma^{MNP}D_N \Psi_{P
i}-\frac{3l^2}{32} {\cal F}_{MN}{\cal F}^{MN}
+\frac{C}{6\sqrt{6}}E^{-1}\epsilon^{MNPQR} {\cal F}_{MN}{\cal
F}_{PQ}{\cal A}_R
\nonumber\\
&&-\frac{3i}{4l}\overline{\Psi}_M^i\Gamma^{MN}\Psi^{N j}
\delta_{ij}
 -\frac{3il}{32}\left( \overline{\Psi}_M^i\Gamma^{MNPQ}
\Psi_{N i}{\cal F}_{PQ}+2\overline{\Psi}^{M i}\Psi^N_i {\cal
F}_{MN}\right)-\frac{6}{l^2}. \label{gaugedfm}
\end{eqnarray}
The supersymmetry transformations at the leading order in spinor
fields read
\begin{eqnarray}
\delta E_M^{~A}&=&\frac{1}{2}\overline{\cal E}^i\Gamma^A\Psi_{M
i},
\nonumber\\
\delta \Psi_{M }^i &=& D_{M}{\cal E}^i+ \frac{il}{16}
\left(\Gamma_{M}^{~NP}-4\delta_{M}^{~N} \Gamma^P \right) {\cal
F}_{NP}{\cal E}^i+\frac{i}{2l}\Gamma_M \delta^{ij}{\cal E}_j,
\nonumber\\
\delta {\cal A}_M &=& \frac{i}{l}\overline{\Psi}_M^i{\cal E}_i.
\label{twostm}
 \end{eqnarray}

To calculate the holographic superconformal anomaly, we need to
expand the fields around the $AdS_5$ vacuum solution
(\ref{metrican}). Geometrically, this  is actually a process of
revealing the asymptotic behaviour of the bulk fields near the
boundary of $AdS_5$ space-time. Correspondingly, the various bulk
symmetries will be reduced to those on the boundary. For examples,
the bulk diffeormorphism invariance of the bulk decomposes into
the diffeomorphism  symmetry on the boundary and the Weyl symmetry
\cite{imbi}, and the the bulk supersymmetry converts into a
superconformal symmetry for an off-shell
 conformal supergravity on the boundary \cite{bala,nish}.

 The procedure of reducing the bulk gauged supergravity to
  the off-shell conformal supergravity on the boundary is
  displayed in a series of works on $AdS_3/CFT_2$,
   $AdS_6/CFT_5$ and $AdS_7/CFT_6$ by Nishimura et al \cite{nish}.
     The key point is  using the equations of motion of the
     bulk fields to
   find their radial coordinate dependence near the boundary
   of $AdS_5$ space. For the spinor field such as
   the gravitino, one should also show how a
  symplectic  Majorana spinor in  five dimensions reduces to
  the chiral spinor on the four-dimensional boundary.
   The reduction from the on-shell
   five-dimensional ${\cal N}=2$ gauged supergravity
   to ${\cal N}=1$ off-shell conformal supergravity
   in four dimensions
   was performed
by Balasubramanian et al. \cite{bala}, so we briefly review their
result and then use it to derive the holographic super-Weyl
anomaly of the supersymmetry current in ${\cal N}=1$
supersymmetric gauge theory.

 As the first step, one should partially fix the local symmetries of bulk
   supergravity
   in the radial direction. According to the $AdS_5$ solution
   (\ref{metrican}), one can choose \cite{bala,nish}
   \begin{eqnarray}
   E_\mu^{~a} (x,r)&=&\frac{l}{r} {e}_\mu^{~a}
    (x)+{\cal O}(r ), ~~
   E_r^{~a}=E_{\mu}^{~\overline{r}}=0,
   ~~E_r^{~\overline{r}}=\frac{l}{r},
   \label{rgf1}
   \end{eqnarray}
   and
   \begin{eqnarray}
   \Psi_r^i(x,r)=0, ~~~ {\cal A}_r (x,r)=0,
   \label{rgf2}
   \end{eqnarray}
to fix the Lorentz symmetry, the supersymmetry and the gauge
symmetry in the $r$-direction, respectively. The gauge-fixing
choice and the torsion-free condition $dE^a+\Omega^a_{~b}\wedge
E^b=0$, further determine the $r$-dependence of the spin
connections,
\begin{eqnarray}
\Omega^{a}_{~\overline{r}}(x,r)&=& -E^{a}(x,r)
=-\frac{l}{r}{e}^{a}(x),\nonumber\\
{\Omega}^{a}_{~b}(x,r) &=&{\omega}^{a}_{~b}(x). \label{redsc}
\end{eqnarray}
In above equations, $\mu,a=0,\cdots,3$ are the Riemannian and
local Lorentz indices on the boundary, respectively, and
$\overline{r}$ is the Lorentz index in the radial direction. We
use the lower case quantities to  denote the boundary values of
bulk fields, i.e., they are independent of the radial coordinate
$r$.

The linearized equation of motion for the gauge field ${\cal
A}_\mu$ near the boundary,
\begin{eqnarray}
E^{-1}\partial_M \left[G^{MN}\partial_N {\cal A}_\mu (x,r)
\right]=0,
\end{eqnarray}
implies that at the leading order in $r$ one can choose ${\cal
A}_\mu (x,r)$  to be independent of $r$,
\begin{eqnarray}
{\cal A}_\mu (x,r) ={A}_\mu (x). \label{redg}
\end{eqnarray}
Furthermore, the linearized equation of motion for the gravitino
is
\begin{eqnarray}
\Gamma^{MNP}D_N \Psi_{P i}+\frac{3i}{2}\Gamma^{MN}
\Psi_{N}^j\delta_{ij}=0.
\end{eqnarray}
The reduced spin connection (\ref{redsc}) and gauge field
(\ref{redg}) as well as the gauge choices (\ref{rgf1}) and
(\ref{rgf2}) lead to the boundary reduction of the bulk covariant
derivative,
 \begin{eqnarray}
 D_r&=&\partial_r, ~~~
 D_\mu
 =\widetilde{D}_\mu (x)-\frac{1}{2r}{\gamma}_\mu \gamma_{5},
 \nonumber\\
 \widetilde{D}_\mu (x) &{\equiv}& \nabla_\mu
 +\frac{1}{4}{\omega}_\mu^{~ab}
 {\gamma}_{~ab}
 +\frac{3}{4}{A}_\mu,
 \label{rederi}
 \end{eqnarray}
 where the four-dimensional convention is defined as the following,
 \begin{eqnarray}
 {\gamma}_{a}=\Gamma_{a},~~ \Gamma_\mu=
 E_\mu^{~a}\Gamma_{a}=\frac{l}{r}
 {\gamma}_{\mu},
 ~~\Gamma^\mu = \overline{E}^\mu_{~a}\Gamma^{a}
 =\frac{r}{l} {\gamma}^{\mu},
 ~~ \gamma_5=\Gamma^{\overline{r}}
 =\Gamma_{\overline{r}},~~\gamma_5^2=1.
 \label{rega}
 \end{eqnarray}
The linearized gravitino equation reduces to
 \begin{eqnarray}
 {\gamma}^{\mu\nu}\left(\partial_r\delta_{ij}-\frac{1}{r}\delta_{ij}
 -\frac{3}{2r}\gamma_5 \epsilon_{ij}\right)\psi_{\nu}^j(x,r)
 -{\gamma}^{\mu\nu\rho}\widetilde{D}_\nu (x)\gamma_5\psi_{\rho i}
 (x,r)=0,
 \end{eqnarray}
Diagonalizing the above equation by combining the two components
of the symplectic Majorana spinor, $\Psi_\mu\equiv \Psi_{\mu
1}+i\Psi_{\mu 2}$, and making the chiral decomposition
$\Psi^R_\mu\equiv \frac{1}{2} (1-\gamma_5)\Psi_\mu$,
 $\Psi^L_\mu \equiv \frac{1}{2} (1+\gamma_5)\Psi_\mu$,
one can  see that near the boundary $r\to 0$, the equation for the
right-handed component reads \cite{bala}
\begin{eqnarray}
\left(\partial_r+\frac{1}{2r}\right)\Psi_\mu^R=0
\end{eqnarray}
and hence the radial dependence of $\Psi_\mu^R$ is
\begin{eqnarray}
\Psi_\mu^R=\left(\frac{2l}{r}\right)^{1/2}{\psi}_\mu^R(x).
\label{redsl}
\end{eqnarray}
The  left-handed  component is not independent, and its radial
coordinate dependent behaviour turns out to be \cite{bala}
\begin{eqnarray}
\Psi_\mu ^L &= & \left(2l\tau\right)^{1/2}{\chi}_\mu^L(x), \nonumber\\
\chi_\mu^L &=&
\frac{1}{3}\gamma^\nu\left(\widetilde{D}_\mu\psi_\nu^R
-\widetilde{D}_\nu\psi_\mu^R\right)-\frac{i}{12}
\epsilon_{\mu\nu}^{~~\lambda\rho} \gamma_5\gamma^\nu
\left(\widetilde{D}_\lambda\psi_\rho^R
-\widetilde{D}_\rho\psi_\lambda^R\right). \label{redsr}
\end{eqnarray}
Therefore, the ${\cal N}=2$  bulk supergravity multiplet
($E_M^{~A},\Psi_M^i, {\cal A}_M$) reduces to the boundary field
(${e}_\mu^{~a}$, ${\psi}_\mu^{R}$, ${A}_\mu$). It is actually the
 8+8 off-shell multiplet of ${\cal N}=1$ conformal supergravity
 since the bulk supersymmetry transformation
(\ref{twostm}) reduces to the one for  ${\cal N}=1$ conformal
supergravity
 \cite{bala}.

The boundary reduction of the bulk supersymmetric transformation
to that for  the conformal supergravity is the following. First,
as done for the bulk gravitino, one must redefine the bulk
supersymmetric transformation parameter, ${\cal E} (x,r)={\cal
E}_1(x,r)+i{\cal E}_2(x,r)$ and decompose it as the chiral
components. The radial coordinate dependence of ${\cal E}^{L,R}$
 should be the same as
as the bulk gravitino,
\begin{eqnarray}
{\cal E}^R(x,r)=\left(\frac{2l}{r}\right)^{1/2}{\epsilon}^R(x),~~~
{\cal E}^L(x,r)= \left(2lr\right)^{1/2} {\eta}^L(x). \label{strp}
\end{eqnarray}
At the leading order in $r$, the bulk supersymmetric
transformations reduces to \cite{bala}
\begin{eqnarray}
\delta {e}_\mu^{~a} &=&-\frac{1}{2}
\overline{\psi}_\mu\gamma^{a}{\epsilon},
\nonumber\\
 \delta {\psi}_\mu
&=&\widetilde{D}_\mu {\epsilon} -\gamma_\mu {\eta} ={\nabla}_\mu
{\epsilon} -\frac{3i}{4}A_\mu \gamma_5 {\epsilon}-\gamma_\mu
{\eta},
\nonumber\\
\delta {A}_\mu  &=& i\left(\overline{\psi}_\mu\gamma_5{\eta}
-\overline{\chi}_\mu \gamma_5\epsilon \right), \label{desut}
\end{eqnarray}
where all the spinorial quantities, $\psi_\mu (x)$, ${\chi}_\mu
(x)$ ${\epsilon} (x)$ and ${\eta}(x)$
 are Majorana spinors constructed from
their chiral components $\psi^R_\mu (x)$, $\chi^L_\mu (x)$,
${\epsilon}^R (x)$ and ${\eta}^L(x)$. Eq.\,(\ref{desut}) shows
that the reduced bulk supersymmetric transformation is indeed the
supersymmetric transformation for ${\cal N}=1$ conformal
supergravity with $\epsilon$ and $\eta$ playing the roles of
parameters for supersymmetry and special supersymmetry
transformations, respectively \cite{kaku,fra}.

\section{Holographic Super-Weyl anomaly of supersymmetry current}

The arising of the super-Weyl anomaly in ${\cal N}=1$
 supersymmetric Yang-Mills
theory from the ${\cal N}=2$ gauged $AdS_5$ supergravity
 lies in two aspects. On one hand,
as a supersymmetric field theory, the supersymmetric variation of
the Lagrangian (\ref{gaugedfm}) of the gauged supergravity is a
total derivative. These total derivative terms cannot be naively
ignored due to the existence of the boundary $AdS_5$. On the other
hand, near the $AdS_5$ boundary, the bulk supersymmetric
transformation  decomposes into the supersymmetry transformation
and the super-Weyl transformation on the boundary. If we require
supersymmetry on the boundary to be preserved, the total
derivative terms should yield the super-Weyl anomaly of the
supersymmetry current via the AdS/CFT correspondence given by
Eq.\,(\ref{acc4}).

 In the following, we shall calculate the supersymmetric
 variation of the gauged ${\cal N}=2$ five-dimensional supergravity
 (\ref{gaugedfm}) and extract out the total derivative terms. Then we shall
 reduce it to the $AdS_5$ boundary and give the holographic supersymmetry
 current anomaly.  However, it should be emphasized that this surface term
 only yields the gauge field background part of the super-Weyl anomaly.
 One must employ a holographic renormalization procedure to reveal
 the gravitational part, just as how the holographic Weyl anomaly
 was found in Ref.\,\cite{hesk}.

 First, the supersymmetric variation of the pure gravitational term
 and the cosmological term gives
 \begin{eqnarray}
 \delta  S_{\rm GR}&=&\delta \int d^5x
  E\left(-\frac{1}{2}{\cal R}-\frac{6}{l^2}\right)
  =\delta \int d^5x
  E\left(-\frac{1}{2}\overline{E}_A^{~M}\overline{E}_B^{~N}
  {\cal R}_{MN}^{~~~AB}
   -\frac{6}{l^2}\right)
  \nonumber\\
  &=&\int d^5x E\left[ -\frac{1}{2}\overline{\cal E}^i\gamma_A\Psi^{M}_i
  \left( {\cal R}_M^{~A}-\frac{1}{2}E_M^{~A} {\cal R}
  -\frac{6}{l^2}E_M^{~A}\right)
  -\nabla_M \left( \overline{E}_A^{~M}\overline{E}_B^{~N}
  \delta\Omega_N^{~AB}\right)\right],
  \label{superv1}
 \end{eqnarray}
where $\overline{E}_A^{~M}$ is the inverse of $E_M^{~A}$.

 The supersymmetry variation of the pure gauge field terms
 (including Chern-Simons term) yields
 \begin{eqnarray}
\delta S_{\rm GA}&=& \int d^5x \delta \left[E \left(-\frac{3
l^2}{32}\right){\cal F}_{MN} {\cal F}^{MN}
 +\frac{C}{6\sqrt{6}} \epsilon^{MNPQR} F_{MN}
 {\cal F}_{PQ} {\cal A}_R\right]\nonumber\\
&=& \int d^5 x \left[-\frac{3l^2}{32}\left(\delta E {\cal F}_{MN}
{\cal F}^{MN}
+4{\cal F}^{MN}\nabla_M\delta {\cal A}_N \right)\right.\nonumber\\
&&\left.+\frac{C}{6\sqrt{6}} \epsilon^{MNPQR} \left(4 \nabla_M
\delta {\cal A}_N
 {\cal F}_{PQ} {\cal A}_R+ {\cal F}_{MN}
 {\cal F}_{PQ}\delta {\cal A}_R\right)\right]\nonumber\\
 &=& \int d^5x \left\{E\left[-\frac{3l^2}{64}\overline{\cal E}^i
 \Gamma^M\Psi_{M i} {\cal F}_{NP} {\cal F}^{NP}\right.\right.\nonumber\\
 &&\left.+\frac{3il}{8}\nabla_M
 \left(F^{MN}\overline{\cal E}^i\Psi_{N i}\right)
 -\frac{3il}{8}\left(\nabla_M {\cal F}^{MN}\right)
 \overline{\cal E}^i\Psi_{N i}
 \right]\nonumber\\
 &&\left. -\frac{iC}{\sqrt{6}l}\epsilon^{MNPQR}
 \left[\frac{2}{3}\nabla_M \left({\cal A}_R
 {\cal F}_{PQ}\overline{\cal E}^i\Psi_{N i}\right)
 +\frac{1}{2}{\cal F}_{MN}{\cal F}_{PQ}\overline{\cal E}^i\Psi_{R i}
   \right]\right\}.
   \label{superv2}
 \end{eqnarray}

As for the supersymmtric variations
 of the terms concerning the gravitino, we start from the
 kinetic terms of the gravitino,
 \begin{eqnarray}
 \delta S_{\rm KT} &=& \delta \int d^5 x \left[-\frac{1}{2} E
 \overline{\Psi}^i_M \overline{E}_A^{~M} \overline{E}_B^{~N}
 \overline{E}_C^{~P}
 \Gamma^{ABC} \left(\nabla_N \Psi_{P i}
 -\frac{3}{4} {\cal A}_N \delta_{ij}\Psi_P^j\right)\right]\nonumber\\
 &=& \int d^5x \left(-\frac{1}{2}\right)\left[ \left(\delta E \right)
 \overline{\Psi}^i_M \Gamma^{MNP} D_N \Psi_{P i}
 +E \left(\delta \overline{\Psi}^i_M\right)
 \Gamma^{MNP} D_N \Psi_{P i}
  \right.\nonumber\\
 && +3 E\left(\delta \overline{E}_A^{~M}\right) \overline{E}_B^{~N}
 \overline{E}_C^{~P}
 \overline{\Psi}_{M}^i\Gamma^{ABC} D_N \Psi_{P i}\nonumber\\
 && \left.+\overline{\Psi}^i_M \Gamma^{MNP}\left(
 \nabla_N \delta \Psi_{P i}-\frac{3}{4} \delta_{ij}
 \left(\delta {\cal A}_N
 \Psi_P^j+ {\cal A}_N\delta \Psi_P^j\right)\right) \right].
 \end{eqnarray}
 Considering only the terms at most quadratic in the
 spinor quantities, we have
 \begin{eqnarray}
\delta S_{\rm KT} &=& \int d^5 x \left(-\frac{1}{2}\right) E
\left\{ \left[\nabla_M \overline{\cal E}^i +\frac{3}{4}{\cal
A}_\mu\delta^{ij}\overline{\cal E}_j-\frac{il}{16} \overline{\cal
E}^i \left(\Gamma_M^{~RS}+4\delta_M^{~R}\Gamma^S
 \right)F_{RS}-
\frac{i}{2l}\delta^{ij}\overline{\cal E}_j\Gamma_M\right]\right.\nonumber\\
&&\times \Gamma^{MNP}\left(\nabla_N\Psi_{P i}
-\frac{3}{4}{\cal A}_N\delta_{ik}\Psi_P^k\right)\nonumber\\
&& + \overline{\Psi}^i_M \Gamma^{MNP}
 \nabla_N \left[\nabla_P {\cal E}_i-\frac{3}{4}{\cal A}_P
 \delta_{ij}{\cal E}^j +\frac{il}{16}
 \left(\Gamma_P^{~RS}
 -4\delta_P^{~R}\Gamma^S \right){\cal E}_i {\cal F}_{RS}
 +\frac{i}{2l}\Gamma_P\delta_{ik}{\cal E}^k\right]\nonumber\\
&&\left.-\frac{3}{4}\delta_{ij}\overline{\Psi}^i_M
\Gamma^{MNP}{\cal A}_N \left[\nabla_P {\cal E}^j+\frac{3}{4}{\cal
A}_P
 \delta^{jk}{\cal E}_k +\frac{il}{16}
 \left(\Gamma_P^{~RS}
 -4\delta_P^{~R}\Gamma^S \right){\cal E}^j {\cal F}_{RS}
 -\frac{i}{2l}\Gamma_P\delta^{jk}{\cal E}_k\right] \right\}
 \nonumber\\
 &=& \int d^5x \left(-\frac{1}{2}\right) E
 \left\{ \left[(\nabla_M \overline{\cal E}^i)\Gamma^{MNP}\nabla_N
 \Psi_{P i}+\overline{\Psi}^i_M\Gamma^{MNP}\nabla_N \nabla_P
 {\cal E}_i\right]\right. ~~~~\bigcirc\hspace{-4mm}1 \nonumber\\
 &&-\frac{3}{4}\left[\delta_{ij}{\cal A}_N \left(\nabla_M
 \overline{\cal E}^i
 \Gamma^{MNP} \Psi_P^j+
 \overline{\Psi}_M^i
 \Gamma^{MNP}\nabla_P{\cal E}^j\right)
 \right.\nonumber\\
 &&\left.-{\cal A}_M\delta^{ij}\overline{\cal E}_j
 \Gamma^{MNP}\nabla_M \Psi_{P i}
 +\delta_{ij}\overline{\Psi}^i_M \Gamma^{MNP}\nabla_N
   ({\cal A}_{P}{\cal E}^j)
  \right] \bigcirc\hspace{-4mm}2\nonumber\\
  && +\frac{il}{16}\left[-{\cal F}_{RS}\overline{\cal E}^i
  \Gamma_M^{~RS}\Gamma^{MNP}\nabla_{N}\Psi_{P i}
  +\overline{\Psi}^i_M \Gamma^{MNP}\nabla_N
   (\Gamma_{P}^{~RS}{\cal F}_{RS}{\cal E}_i)
    \right]~~~~ \bigcirc\hspace{-4mm}3 \nonumber\\
  &&+\frac{il}{16}\frac{3}{4}{\cal A}_N {\cal F}_{RS}\delta_{ij}
  \left(\overline{\cal E}^i\Gamma_{M}^{~RS}\Gamma^{MNP}
  \Psi_{P}^j
  -\overline{\Psi}^i_M\Gamma^{MNP}\Gamma_{P}^{~RS}
  {\cal E}^j  \right)~~~~ \bigcirc\hspace{-4mm}4 \nonumber\\
  &&-\frac{il}{4}\left[{\cal F}_{MR}
  \overline{\cal E}^i\Gamma^{R}\Gamma^{MNP}\nabla_N
  \Psi_{P i}+\overline{\Psi}^i_M\Gamma^{MNP}\nabla_N\left(
  \Gamma^{R} F_{PR}{\cal E}_i\right)\right]~~~~
  \bigcirc\hspace{-4mm}5 \nonumber\\
  &&+\frac{il}{4} \frac{3}{4}\delta_{ij} {\cal A}_N
  \left( {\cal F}_{MR}
  \overline{\cal E}^i\Gamma^{R}\Gamma^{MNP}
  \Psi_{P}^j+{\cal F}_{PR}\overline{\Psi}^i_M\Gamma^{MNP}
  \Gamma^{R}{\cal E}^j\right)~~~~ \bigcirc\hspace{-4mm}6 \nonumber\\
  &-&\frac{i}{2l}\left[\delta^{ij}
  \overline{\cal E}_j\Gamma_{\mu}\Gamma^{MNP}\nabla_N
  \Psi_{P i}-\delta_{ij}\overline{\Psi}^i_M\Gamma^{MNP}
  \nabla_N\left(
  \Gamma_P{\cal E}^j\right)\right]
  ~~~~ \bigcirc\hspace{-4mm}7 \nonumber\\
  &&\left.+\frac{3}{4}\frac{i}{2l}{\cal A}_N \delta^i_{~j}\left(
  \overline{\cal E}_i\Gamma_{M}\Gamma^{MNP}
  \Psi_{P}^j+\overline{\Psi}^j_M\Gamma^{MNP}
  \Gamma_P{\cal E}_i\right)
  \right\} ~~~~ \bigcirc\hspace{-4mm}8.
  \label{superv3}
 \end{eqnarray}
 We list the calculation results in the following:
 \begin{eqnarray}
 \bigcirc\hspace{-3mm}1\,&=&\nabla_M
 \left(\overline{\cal E}^i\Gamma^{MNP}\nabla_N
 \Psi_{P i}\right)-\overline{\cal E}^i
 [\nabla_M, \Gamma^{MNP}]\nabla_N
 \Psi_{P i}\nonumber\\
 && -\frac{1}{8}R_{MN AB}
 \overline{\cal E}^i\left(\Gamma^{MNP}\Gamma^{AB}
 +\Gamma^{AB}\Gamma^{MNP}\right)\Psi_{P i}\nonumber\\
 &=& \nabla_M
 \left(\overline{\cal E}^i\Gamma^{MNP}\nabla_N
 \Psi_{P i}\right)
 +\left( R_{M}^{~N}-\frac{1}{2}R\delta_{M}^{~N}\right)
 \overline{\cal E}^i \Gamma^{M}\Psi_{N i}; \nonumber \\
 \bigcirc\hspace{-3mm}2\,
 &=& -\frac{3}{4}\nabla_M \left(\delta^{ij} {\cal A}_N
 \overline{\cal E}_i\Gamma^{MNP}\Psi_{P j}\right)
 +\frac{3}{4}\overline{\cal E}^i\Gamma^{MNP}\Psi_M^j \delta_{ij}
 {\cal F}_{NP}; \nonumber \\
 \bigcirc\hspace{-3mm}3\,&=&\frac{il}{16}
 \nabla_M\left({\cal F}_{RS}\overline{\cal E}^i\Gamma_P^{~RS}
 \Gamma^{MNP}\Psi_{N i}\right)
 -\frac{il}{8}{\cal F}_{RS}\overline{\cal E}^i\Gamma^{NPRS}
 \left(\nabla_N \Psi_{P}\right)_i+
 \frac{3il}{4}{\cal F}^{MN}\overline{\cal E}^i \left(\nabla_M
 \Psi_{N}\right)_i;
\nonumber \\
    \bigcirc\hspace{-3mm}4\,&=&\frac{3il}{32}
    {\cal F}_{RS}{\cal A}_M \delta_{ij}\overline{\cal E}^i
    \Gamma^{MNRS}\Psi_N^j
    -\frac{9il}{16}\delta_{ij}{\cal F}^{MN}
    {\cal A}_M \overline{\cal E}^i\Psi_N^j
    ; \nonumber \\
 \bigcirc\hspace{-3mm}5\, &=&  \frac{il}{4}
 \nabla_M \left( \overline{\cal E}^i\Gamma^S\Gamma^{MNP}
 \Psi_N^i {\cal F}_{PS}\right)
 +\frac{il}{2} {\cal F}_{MR}
 \overline{\cal E}^i\Gamma^{MNPR}
 \left(\nabla_N\Psi_P\right)_i;
 \nonumber \\
    \bigcirc\hspace{-3mm}6\,&=&
   -\frac{3il}{8}
{\cal F}_{MR}{\cal A}_N\delta_{ij} \overline{\cal
E}^i\Gamma^{MNPR}
\Psi_P^j; \nonumber \\
     \bigcirc\hspace{-3mm}7\,&=&
     \frac{i}{2l}\delta_{ij}\left[-3\overline{\cal E}^i
     \Gamma^{MN}\nabla_M\Psi_{N}^j+\overline{\Psi}_M^i
    \left[\Gamma^{MNP},\nabla_N\right]\left(\Gamma_P{\cal E}^j\right)
    +3 \overline{\Psi}_M^i\nabla_N\left(\Gamma^{MN} {\cal E}^j\right)
    \right]
\nonumber\\
&=&\frac{3i}{2l}\nabla_M \left( \overline{\cal E}^i
\Gamma^{MN}\Psi_{N}^j\right)
 -\frac{3i}{l}\delta^{ij}\overline{\cal E}^i\Gamma^{MN}
 \nabla_M\Psi_{N}^j ;\nonumber\\
  \bigcirc\hspace{-3mm}8\,&=&-\frac{9i}{4l}{\cal A}_M \overline{\cal E}^i
  \Gamma^{MN}\Psi_{N i}.
  \label{superv4}
 \end{eqnarray}
 The supersymmetric variation of the gravitino mass term is
 \begin{eqnarray}
 \delta S_{\rm GM} &=& \delta \left[\frac{3i}{4l}
 \int E \overline{\Psi}_M^i \Gamma^{MN}
 \Psi_N^j \delta_{ij} \right]
 =\frac{3i}{4l}\int d^5x E\left[
 \left(\delta \overline{\Psi}_M^i\right) \Gamma^{MN}\Psi_N^j
 +\overline{\Psi}_M^i \Gamma^{MN}\delta \Psi_N^j \right]\delta_{ij}
  \nonumber\\
 &=&\frac{3i}{4l}\int d^5x e\left\{\left[\nabla_M \overline{\cal E}^a
+\frac{3}{4}{\cal A}_M\delta^{ik}\overline{\cal E}_k-\frac{il}{16}
\overline{\cal E}^i \left(\Gamma_M^{~PQ}+4\delta_M^{~P}\Gamma^Q
 \right)F_{PQ}\right.\right.\nonumber\\
 &&\left.-
\frac{i}{2l}\delta^{ac}\overline{\cal E}_k\Gamma_M\right]
\Gamma^{MN} \Psi_N^b\nonumber\\
&& \left.+\overline{\Psi}_M^i \Gamma^{MN} \left[\nabla_N {\cal
E}^j+\frac{3}{4}{\cal A}_N
 \delta^{jk}{\cal E}_k +\frac{il}{16}
 \left(\Gamma_N^{~}
 -4\delta_N^{~P}\Gamma^Q \right){\cal E}^j {\cal F}_{PQ}
 -\frac{i}{2l}\Gamma_N\delta^{jk}{\cal E}_k\right]\right\}\delta_{ij}
 \nonumber\\
 &=&\frac{3i}{4l}\int d^5x E\left\{\delta_{ij}\left[
 (\nabla_M \overline{\cal E}^i)\Gamma^{MN}\Psi_N^j
  + \overline{\Psi}^i_M\Gamma^{MN}\nabla_N{\cal E}^j\right]
  \right.
 \nonumber\\
 && +\frac{3}{4}\delta^i_{~j}\left({\cal A}_M \overline{\cal E}_i
 \Gamma^{MN}\Psi_N^j+{\cal A}_N \overline{\Psi}_M^j\Gamma^{MN}
 {\cal E}_i \right) \nonumber\\
 && -\frac{il}{16}\delta_{ij} {\cal F}_{RP}\left(
 \overline{\cal E}^i\Gamma_M^{~RP}
 \Gamma^{MN}\Psi_N^j
  -\overline{\Psi}^i_M\Gamma^{MN}\Gamma_N^{~RP}
 {\cal E}^j \right)\nonumber\\
 &&-\frac{il}{4}\delta_{ij}\left(
 \overline{\cal E}^i\Gamma^{P}
 \Gamma^{MN}\Psi_N^j {\cal F}_{MP}
  +\overline{\Psi}^i_M\Gamma^{MN}\Gamma^{P}{\cal F}_{NP}
 {\cal E}^j \right)\nonumber\\
 &&-\frac{2i}{l}\delta^i_{~j}
 \left(\overline{\cal E}_i\Gamma^{M}\Psi_{M}^j
 +\overline{\Psi}^j_M\Gamma^M{\cal E}_i\right)\nonumber\\
&=&\int d^5x E \left[\frac{3i}{2l}
 \left(\nabla_M \overline{\cal E}^i\right)
 \Gamma^{MN}\Psi_N^j \delta_{ij}
 -\frac{9i}{8}{\cal A}_M \overline{\cal E}^i
 \Gamma^{MN}\Psi_{N i}
 \right.\nonumber\\
 &&\left.
 +\frac{3}{16}\delta_{ij}{\cal F}^{MN}\overline{\cal E}^i\Gamma_M\Psi_N^j
 -\frac{3}{16}\delta_{ij}{\cal F}_{MN}
 \overline{\cal E}^i\Gamma^{MNP} \Psi_P^j
 -\frac{3}{l^2}\overline{\cal E}^i\Gamma^M\Psi_{M i}
 \right].
 \label{superv5}
 \end{eqnarray}
 Finally, the supersymmetric variation of the interaction terms between
 the gravitino $\Psi_M$ and the graviphoton $A_M$ gives
 \begin{eqnarray}
 &&\delta \int d^5 x E \left(-\frac{3il}{32}\right)\left(
 \overline{\Psi}_M^i \Gamma^{MNPQ}\Psi_{N i} {\cal F}_{PQ}
 +2 \overline{\Psi}_M^i \Psi_i^N {\cal F}_{MN}\right)\nonumber\\
 &=& -\frac{3il}{32}\int d^5x E  \left\{ {\cal F}_{PQ} \left(
 \left[\nabla_M \overline{\cal E}^i
+\frac{3}{4}{\cal A}_M\delta^{ik}\overline{\cal E}_k-\frac{il}{16}
\overline{\cal E}^i \left(\Gamma_M^{~RS}+4\delta_M^{~R}\Gamma^S
 \right){\cal F}_{RS}\right.\right.\right.\nonumber\\
&&\left.- \frac{i}{2l}\delta^{ik}\overline{\cal
E}_k\Gamma_M\right]
\Gamma^{MNPQ}\Psi_{N i}\nonumber\\
&&\left.+\overline{\Psi}_M^i \Gamma^{MNPQ} \left[\nabla_Q {\cal
E}_i-\frac{3}{4}{\cal A}_Q
 \delta_{ij}{\cal E}^j +\frac{il}{16}
 \left(\Gamma_Q^{~RS}
 -4\delta_Q^{~R}\Gamma^S \right){\cal E}_i {\cal F}_{RS}
 +\frac{i}{2l}\Gamma_Q\delta_{ik}{\cal E}^k   \right]\right)\nonumber\\
&& +2 {\cal F}^{MN}\left(\left[\nabla_M \overline{\cal E}^i
+\frac{3}{4}{\cal A}_M\delta^{ik}\overline{\cal E}_k-\frac{il}{16}
\overline{\cal E}^i \left(\Gamma_M^{~RS}+4\delta_M^{~R}\Gamma^S
 \right)F_{RS}-
\frac{i}{2l}\delta^{ik}\overline{\cal E}_k\Gamma_M \right]
\Psi_{N i}\right.\nonumber\\
&& \left.+\overline{\Psi}_{M}^i \left[\nabla_N {\cal
E}_i-\frac{3}{4} {\cal A}_N
 \delta_{ij}{\cal E}^j +\frac{il}{16}
 \left(\Gamma_N^{~RS}
 -4\delta_N^{~R}\Gamma^S \right){\cal E}_i {\cal F}_{RS}
 +\frac{i}{2l}\Gamma_N\delta_{ik}{\cal E}^k   \right]\right)\nonumber\\
&&\left. +\cdots \right\}\nonumber\\
&=& -\frac{3il}{32}\int d^5x E\left\{ 2\left[ F^{MN} \left(
\nabla_M\overline{\cal E}^i \Psi_{N i}
+\overline{\Psi}_M^i\nabla_N{\cal E}_i\right)
+\frac{3}{4}\delta^{ij}{\cal A}_M {\cal F}^{MN}\left(
\overline{\cal E}_i \Psi_{N j}
+\overline{\Psi}_{N i} {\cal E}_j \right)\right.\right.\nonumber\\
&&+\frac{il}{16}{\cal F}^{MN}{\cal F}^{PQ}\left( -\overline{\cal
E}^i \Gamma_{MPQ}\Psi_{N i} +\overline{\Psi}^i_M \Gamma_{NPQ}{\cal
E}_{i}\right) -\frac{il}{4}{\cal F}^{MN}\left( \overline{\cal E}^i
\Gamma^{P}\Psi_{N i}{\cal F}_{MP} +\overline{\Psi}^i_M
\Gamma^{P}{\cal E}_{i}{\cal F}_{NP}\right)
\nonumber\\
&&\left.-\frac{i}{2l}{\cal F}^{MN} \left(
\delta^{ij}\overline{\cal E}_i \Gamma_{M}\Psi_{N j}
-\delta_{ij}\overline{\Psi}^i_M \Gamma_{N}{\cal E}^{j}\right)
\right]
\nonumber\\
&&+{\cal F}_{PQ}\left[ \left(\nabla_M\overline{\cal E}^i
\Gamma^{MNPQ}\Psi_{Ni} +\overline{\Psi}_M^i
\Gamma^{MNPQ}\nabla_N{\cal E}_{i}\right)\right.
\nonumber\\
&& +\frac{3}{4}\left({\cal A}_M\delta^{ij}\overline{\cal E}_i
\Gamma^{MNPQ}{\Psi}_{N j} -{\cal
A}_N\delta_{ij}\overline{\Psi}^i_M
\Gamma^{MNPQ}{\cal E}^{j}\right)\nonumber\\
&&-\frac{il}{16}{\cal F}_{RS} \left(\overline{\cal
E}^i\Gamma_M^{~RS} \Gamma^{MNPQ}\Psi_{N i} -\overline{\Psi}^i_M
\Gamma^{MNPQ}
\Gamma_N^{~RS}{\cal E}_{i}\right)\nonumber\\
&& -\frac{il}{4}\left( \overline{\cal E}^i\Gamma^{R}
\Gamma^{MNPQ}\Psi_{N i}{\cal F}_{MR} +\overline{\Psi}^i_M
\Gamma^{MNPQ}
\Gamma^{R}{\cal E}_{i} {\cal F}_{NR} \right)\nonumber\\
&&\left.\left.-\frac{i}{2l}\left(\delta^{ij}\overline{\cal
E}_i\Gamma_M \Gamma^{MNPQ}\overline{\Psi}_{N j}
-\delta_{ij}\overline{\Psi}^i_M \Gamma^{MNPQ}\Gamma_N {\cal
E}^{j}\right)\right]
 \right\}\nonumber\\
&=& \int d^5x E\left[-\frac{3il}{8} \left(\nabla_M \overline{\cal
E}\right)^i\Psi_{N i}{\cal F}^{MN} -\frac{9il}{32}\delta^{ij}{\cal
A}_M {\cal F}^{MN} \overline{\cal E}_i \Psi_{N j}
\right.\nonumber\\
&& -\frac{3il}{16} \left(\nabla_M \overline{\cal
E}\right)^i\Gamma^{MNPQ} \Psi_{N i}{\cal
F}_{PQ}-\frac{9il}{64}{\cal  A}_M {\cal F}_{PQ}\delta^{ij}
\overline{\cal E}_i\Gamma^{MNPQ}\Psi_{Nj}\nonumber\\
&&+\frac{3l^2}{64}{\cal F}^{MN}{\cal F}_{MN}\overline{\cal
E}^i\Gamma^P \Psi_{P i}+\frac{3l^2}{64}E^{-1} \epsilon^{MNPQR}
{\cal F}_{MN}{\cal F}_{PQ} \overline{\cal E}^i
\Psi_{R i}\nonumber\\
&& +\frac{3l^2}{32}\overline{\cal E}^i\Gamma^{Q}
 \Psi_{N a}{\cal F}^{NP}{\cal F}_{PQ}
 -\frac{3l^2}{32}\overline{\cal E}^i\Gamma_{P}
 \Psi_{N i}{\cal F}^{MP}{\cal F}_{MN}\nonumber\\
&& \left.-\frac{3}{16}\delta^{ij}\overline{\cal E}_i\Gamma_M
\Psi_{Nj} {\cal F}^{MN}+\frac{3}{16}\delta^{ij} \overline{\cal
E}_i\Gamma^{MNP} \Psi_{Mj} {\cal F}_{NP}\right]. \label{superv6}
 \end{eqnarray}
 Putting the above supersymmetric variations (\ref{superv1})
--- (\ref{superv6}) together, we get
\begin{eqnarray}
\delta S &=& \int d^5 x E \nabla_M \left(
%-e^\mu_{~r}e^\nu_{~s}
%\delta \omega_{\nu}^{~rs}
-\frac{9il}{16}
 \overline{\cal E}^i\Psi_{N i}{\cal F}^{MN}
 -\frac{1}{2}\overline{\cal E}^i\Gamma^{MNP}\nabla_N
 \Psi_{P i}+ \frac{3}{8}
  \overline{\cal E}^i\Gamma^{MNP}
 \Psi_{P}^j\delta_{ij}{\cal A}_N \right.\nonumber\\
 &&\left. -\frac{3il}{32}
 \overline{\cal E}^i\Gamma^{MNPQ}
 \Psi_{N i}{\cal F}_{PQ}
 +\frac{9}{4}\overline{\cal E}^i\Gamma^{MN}
 \Psi_{N}^j\delta_{ij}
 +\frac{2iC}{3\sqrt{6}l}E^{-1}
 \epsilon^{MNPQR}\overline{\cal E}^i \Psi_{R}{\cal A}_N
 {\cal F}_{PQ}\right)
 \nonumber\\
  &&
  +\int d^5x \left(-\frac{iC}{2\sqrt{6}l}+\frac{3l^2}{64}\right)
  \epsilon^{MNPQR}\overline{\cal E}^i
  \Psi_{R}{\cal F}_{MN}{\cal F}_{PQ}.
  \label{var1}
\end{eqnarray}

 In above calculation, we have made use of the following
 relations,
 \begin{eqnarray}
 \Psi^i &=& C^{-1}\Omega^{ij}\overline{\Psi}_j^T=C^{-1}\overline{\Psi}^{iT},
 ~~~\overline{\Psi}^i=-\Psi^{iT}C,\nonumber\\
&&   \overline{\Psi}^i\Gamma_{M_1\cdots M_n}\Phi_i =
 -\Psi^{iT}C\Gamma_{M_1\cdots M_n}C^{-1}\overline{\Phi}_i^T\nonumber\\
&=&\left\{\begin{array}{l}
 -\Psi^{iT}\Gamma_{M_1\cdots M_n}\overline{\Phi}^T_i
=\overline{\Phi}_i\Gamma_{M_1\cdots M_n} \Psi^i
=-\overline{\Phi}^i\Gamma_{M_1\cdots M_n} \Psi_i,~~ n=0,1,4,5,\\
\Psi^{iT}\Gamma_{M_1\cdots M_n}\overline{\Phi}^T_i
=-\overline{\Phi}_i\Gamma_{M_1\cdots M_n} \Psi^i
=\overline{\Phi}^i\Gamma_{M_1\cdots M_n} \Psi_i,~~ n=2,3,
\end{array}
\right. , \nonumber\\
\Gamma_{MN}&=&\frac{1}{2}[\Gamma_M,\Gamma_N], ~~
\Gamma^{MNP}=-\frac{1}{2!}E^{-1}\,
\epsilon^{MNPQR}\Gamma_{QR},\nonumber\\
\Gamma^{MNPQ}&=& E^{-1}\, \epsilon^{MNPQR}\Gamma_{R},
~~\Gamma_{MNPQR}=E\, \epsilon_{MNPQR}.
\nonumber \\
 \Gamma_{MN}\Gamma_{PQ}
&=& E\,\epsilon_{MNPQR}\Gamma^R- \left(G_{MP}
G_{NQ}-G_{MQ}G_{NP}\right),
\nonumber\\
\Gamma_{M}\Gamma_{NP} &=& \Gamma_{MNP}
+G_{MN}\Gamma_P-G_{MP}\Gamma_N, \nonumber\\
\Gamma^{MNP}\nabla_N \nabla_P \Psi_i &=&
\frac{1}{2}\Gamma^{MNP}[\nabla_N, \nabla_P] \Psi_i =
\frac{1}{8}\Gamma^{MNP}{\cal R}_{NP AB}\Gamma^{AB}\Psi_i.
 \end{eqnarray}
 Due to the nocommutativity between $\nabla_M$ and
 $\Gamma_{M_1\cdots M_n}$, we reiteratively use the following operations,
 \begin{eqnarray}
 \Gamma_{M_1\cdots M_n} \nabla _M (\cdots)
 &=&\left[\Gamma_{M_1\cdots M_n},\nabla _M\right] (\cdots)
 +\nabla_M \left[\Gamma_{M_1\cdots M_n} (\cdots)\right].
 \end{eqnarray}
 It is convenient to choose the inertial coordinate system,
 i.e. the Christoffel symbol $\Gamma^{M}_{~NP}=0$. Consequently,
 the metricity condition leads to
 \begin{eqnarray}
 \partial_M E_N^{~A}=0,
 \end{eqnarray}
 and hence the modified spin connection,
 \begin{eqnarray}
 \Omega_{M AB}&=&\frac{1}{2}\overline{E}_A^{~N}\left(\partial_M E_{N B}
 -\partial_N E_{M B}\right)
 -\frac{1}{2}\overline{E}_B^{~N}\left(\partial_M E_{N A}
 -\partial_N E_{M A}\right)\nonumber\\
 && -\frac{1}{2}\overline{E}_A^{~P}
 \overline{E}_B^{~Q}\left(\partial_P E_{Q c}
 -\partial_Q E_{PC}\right)E_{M}^{~C}
 +\frac{1}{4}\left(\overline{\Psi}_M\Gamma_A\Psi_B+
 \overline{\Psi}_A\Gamma_M\Psi_B-\overline{\Psi}_M\Gamma_B\Psi_A
  \right),
 \end{eqnarray}
  keeps only the quadratic fermionic terms.
 We have also considered the Ricci and Bianchi identities for
 the Riemannian curvature tensor and the Abelian gauge field
 \begin{eqnarray}
 \epsilon^{MNPQR} {\cal R}_{SPQR}=0, ~~~
 \epsilon^{MNPQR}\nabla_N {\cal R}_{STPQ}=0,~~~
 \epsilon^{MNPQR}\nabla_N {\cal F}_{QR}=0.
 \end{eqnarray}

If we choose the indefinite Chern-Simons coefficient in
Eq.\,(\ref{var1}) as
\begin{eqnarray}
C=-\frac{3i\sqrt{6}l^3}{32}, \label{var2}
\end{eqnarray}
the above supersymmetric variation of ${\cal N}=2$ gauged
supergravity action in five dimensions is a total derivative,
\begin{eqnarray}
\delta S &=& \int d^4x \int_{r=\epsilon} dr  \partial_M
\left[E\left(
%-e^\mu_{~r}e^\nu_{~s}\delta \omega_{\nu}^{~rs}
-\frac{9il}{16}
 \overline{\cal E}^i\Psi_{N i}{\cal F}^{MN}
 -\frac{1}{2}\overline{\cal E}^i\Gamma^{MNP}\nabla_N
 \Psi_{P i}+ \frac{3}{8}
  \overline{\cal E}^i\Gamma^{MNP}
 \Psi_{P}^j\delta_{ij}{\cal A}_N \right.\right.\nonumber\\
 &-&\left. \left.\frac{3il}{32}E^{-1}\epsilon^{MNPQR}
 \overline{\cal E}^i\Gamma_{R}
 \Psi_{N i}{\cal F}_{PQ}
 +\frac{9}{4}\overline{\cal E}^i\Gamma^{MN}
 \Psi_{N}^i\delta_{ij}
 +\frac{l^2}{16}E^{-1}
 \epsilon^{MNPQR}
 \overline{\cal E}^i \Psi_{R}{\cal A}_N {\cal F}_{PQ}\right)\right],
\label{var3}
\end{eqnarray}
which is the typical feature of a supersymmetric field theory.

 Now we can extract out the holographic super-Weyl anomaly from above
 total derivative
 terms as Witten \cite{witt1} did in finding out the chiral
 $R$-symmetry anomaly. Considering the $r$-dependence of bulk fields
 and of the supersymmetry transformation parameter ${\cal E}^i$ given in
 (\ref{rgf1}), (\ref{rgf2}), (\ref{redsc}), (\ref{redg}), (\ref{rega}),
 (\ref{redsl}), (\ref{redsr}) and (\ref{strp}), and taking
 the boundary limit $\epsilon\to 0$ after we integrate over
 the radial coordinate to the near boundary cut-off $r=\epsilon$,
 one can see that the non-vanishing contribution comes only from the
  term of the form $E^{-1}\epsilon^{MNPQR}\overline{\cal E}^i\Gamma_{R}
 \Psi_{N i}{\cal F}_{PQ}$. Thus we obtain
 \begin{eqnarray}
 \delta S &=& \frac{3il^3}{32}\int d^4 x \epsilon^{\mu\nu\lambda\rho}
 F_{\nu\lambda}\overline{\eta}\gamma_\rho\chi_\mu.
 \label{fvar}
  \end{eqnarray}
 In deriving Eq.\,(\ref{fvar}) we have used  the fact  that
 the metric on the boundary should be the induced metric
 \begin{eqnarray}
 \widetilde{g}_{\mu\nu}(x,r)=\frac{l^2}{r^2}{g}_{\mu\nu}(x,r)
\end{eqnarray}
rather than ${g}_{\mu\nu}(x,r)$ \cite{bian1}. Eq.\,(\ref{fvar})
will lead to the super-Weyl anomaly in terms of the AdS/CFT
correspondence since it is proportional to the special
supersymmetry transformation parameter $\eta$. Considering
Eq.\,(\ref{redsr}), we have
 \begin{eqnarray}
 \delta S &=& \int d^4x \overline{\eta}\gamma^\mu s_\mu \nonumber\\
 &=& \frac{3il^3}{32}\int d^4 x \epsilon^{\mu\nu\lambda\rho}
 F_{\nu\lambda}\overline{\eta}\gamma_5\gamma_\rho \gamma^\alpha
 \left[ \frac{1}{3} \left(D_\mu\psi_\alpha- D_\mu\psi_\alpha\right)
 -\frac{i}{6}\epsilon_{\mu\alpha\sigma\delta}\gamma_5 D^\sigma\psi_\delta
  \right]\nonumber\\
  &=& -\frac{l^3}{16}\int d^4x \left[F^{\mu\nu}D_\mu \psi_\nu
  +\epsilon^{\mu\nu\lambda\rho}\gamma_5 F_{\mu\nu} D_\lambda\psi_\rho
  +\frac{1}{2}\sigma^{\mu\nu} F_{\nu\lambda}
  \left(D_\mu\psi^\lambda-D^\lambda\psi_\mu\right)\right],
 \end{eqnarray}
 where we have used the $\gamma$-matrix algebraic relations,
 \begin{eqnarray}
 \gamma^\mu\gamma^\nu=g^{\mu\nu}-i\gamma^{\mu\nu},
 ~~~~\gamma_5\gamma^{\mu\nu}=\frac{i}{2}\epsilon^{\mu\nu\lambda\rho}
 \gamma_{\lambda\rho}.
 \end{eqnarray}

If we switch on the overall gravitational constant factor
$-1/(8\pi G^{(5)})$ on the classical Lagrangian (\ref{gaugedfm}),
and consider the following relations among the $AdS_5$ radius $l$,
string coupling $g_s$, the number $N$ of the $D3$-branes, the
five- and ten-dimensional gravitational constants connected  by
the compactification on $S^5$ of radius $l$,
\begin{eqnarray}
G^{(5)}=\frac{G^{(10)}}{\mbox{Volume}\,
(S^5)}=\frac{G^{(10)}}{l^5\pi^3}, ~~~ G^{(10)}=8\pi^6g^2_s,
~~~l=\left(4\pi N g_s\right)^{1/4},
\end{eqnarray}
Eq.\,(\ref{fvar}) yields the gauge field part of the holographic
super-Weyl anomaly,
\begin{eqnarray}
\gamma_\mu s^\mu &=& \frac{1}{8\pi G^{(5)}}\frac{l^3}{16}
\left[F^{\mu\nu}D_\mu \psi_\nu
  +\epsilon^{\mu\nu\lambda\rho}\gamma_5 F_{\mu\nu} D_\lambda\psi_\rho
  +\frac{1}{2}\sigma^{\mu\nu} F_{\nu\lambda}
  \left(D_\mu\psi^\lambda-D^\lambda\psi_\mu\right)\right]\nonumber\\
&=& \frac{N^2}{64\pi^2}\left[F^{\mu\nu}D_\mu \psi_\nu
  +\epsilon^{\mu\nu\lambda\rho}\gamma_5 F_{\mu\nu} D_\lambda\psi_\rho
  +\frac{1}{2}\sigma^{\mu\nu} F_{\nu\lambda}
  \left(D_\mu\psi^\lambda-D^\lambda\psi_\mu\right)\right].
\label{gtra}
\end{eqnarray}
This is the super-Weyl anomaly of supersymmetry current
contributed from the external gauge field background at the
leading order of large-$N$ expansion of ${\cal N}=1$ $SU(N)$
supersymmetric gauge theory.

\section{Summary and Discussion}

We have investigated the super-Weyl anomaly of ${\cal N}=1$
supersymmetric Yang-Mills theory in the external ${\cal N}=1$
conformal supergravity background via the AdS/CFT correspondence.
With the speculation that at low-energy the type IIB supergravity
in $AdS_5\times X^5$ background should reduce to the gauged
supergravity in five dimensions since such a background provides a
spontaneous compactification on $X^5$, there should exists a
holographic correspondence between ${\cal N}=2$ conformal
supergravity in five dimensions   and ${\cal N}=1$ supersymmetric
Yang-Mills theory at the fixed point of its renormalization group
flow. The five-dimensional ${\cal N}=2$ gauged  supergravity  has
an $AdS_5$ classical solution  which preserves the full
supersymmetry. Around this $AdS_5$ vacuum configuration, the
five-dimensional
 ${\cal N}=2$ on-shell gauged supergravity multiplet reduces to the
${\cal N}=1$ off-shell conformal supergravity mutiplet on the
boundary of $AdS_5$ space. Correspondingly, the bulk ${\cal N}=2$
supersymmetry transformation converts into the ${\cal N}=1$
superconformal transformation in four dimensions, which consists
of the  supersymmetry transformation for ${\cal N}=1$ Poincar\'{e}
supergravity and super-Weyl transformation. With these facts in
mind, we calculate the supersymmetry variation of the ${\cal N}=2$
gauged supergravity in five dimensions and obtain the total
derivative terms. Further, we reduce the total derivative terms to
the boundary of $AdS_5$ space using the boundary reduction of the
bulk fields. Considering the incompatibility of the ${\cal N}=1$
Poincar\'{e} supersymmetry and the super-Weyl symmetry, we extract
out the super-Weyl anomaly of ${\cal N}=1$ supersymmetric gauge
theory.

   However, as shown in Eq.\,(\ref{gtra}), we only reveal
   the contribution from the external gauged field when ${\cal N}=1$
supersymmetric gauge theory couples to ${\cal N}=1$ conformal
supergravity background. As shown  in Ref.\cite{abb},
 there usually should also has a contribution relevant to
 the external gravitational  background,
\begin{eqnarray}
\gamma_\mu s^{\mu} \sim \frac{N^2}{128\pi^2}
R^{\mu\nu\lambda\rho}\gamma_{\mu\nu} D_\lambda\psi_\rho.
\label{gc}
\end{eqnarray}
The reason for not having revealed the contribution of the
gravitational background field is not yet fully clear. However,
since the super-Weyl anomaly shares a supermultiplet with the Weyl
anomaly and the chiral R-symmetry anomaly, we have the following
two speculations based on the process of deriving the
gravitational background parts in both the holographic Weyl and
chiral anomalies \cite{ahar,blau}.

First, it is possible that this term cannot be revealed within the
five-dimensional ${\cal N}=2$ gauged supergravity itself, and one
must consider the supersymmetric action consisting of
 higher-order gravitational terms such as the Gauss-Bonnet term.
As given in Ref.\,\cite{grisaru}, for a general ${\cal N}=1$
supersymmetric gauge theory with $N_v$ vector and $N_\chi$ chiral
multiplets in an external supergravity background, the chiral
R-symmetry anomaly and the Weyl anomaly take the forms,
\begin{eqnarray}
\nabla_\mu j^{(5)\mu} &=&\frac{c-a}{24\pi^2} R_{\mu\nu\lambda\rho}
\widetilde{R}^{\mu\nu\lambda\rho}+\frac{5a-3c}{9\pi^2}F_{\mu\nu}
\widetilde{F}_{\mu\nu}, \nonumber\\
\theta^\mu_{~\mu} &=& \frac{c}{16\pi^2}C_{\mu\nu\lambda\rho}
C^{\mu\nu\lambda\rho}
-\frac{a}{16\pi^2}\widetilde{R}_{\mu\nu\lambda\rho}
\widetilde{R}^{\mu\nu\lambda\rho}+\frac{c}{6\pi^2}
F_{\mu\nu}F^{\mu\nu}.
\end{eqnarray}
The coefficients are purely determined by the field contents. For
a supersymmetric theory in the weak coupling limit, they are
\cite{grisaru}
 \begin{eqnarray}
 c=\frac{1}{24}\left(3 N_v+N_{\chi}\right), ~~~
 a=\frac{1}{48}\left(9 N_v+N_{\chi}\right).
 \label{cc}
 \end{eqnarray}
 In the pioneering work by Witten \cite{witt1},
 the chiral $R$-symmetry anomaly of ${\cal N}=4$ $SU(N)$
 SYM in the large-$N$ limit is perfectly
 reproduced from the Chern-Simons (CS) term in ${\cal N}=8$
 five-dimensional $SO(6)$ gauged supergravity. The reason is that
 the field contents of ${\cal N}=4$ SYM
 can be considered as one ${\cal N}=1$ vector multiplet
 plus three chiral multiplets in the adjoint representation
 of $SU(N)$, and  this leads to $c=a=1/4$ \cite{blau}. Thus
 there is no anomalous term  relevant to the gravitational background
 field in the chiral anomaly and the CS term is
 fully responsible for the holographic
 origin of the chiral R-symmetry anomaly. This
 also explains the absence of the
 $R_{\mu\nu\lambda\rho}R^{\mu\nu\lambda\rho}$ term
 in the holographic Weyl
 anomaly of ${\cal N}=4$ SYM found in Ref.\,\cite{hesk}.
 For the ${\cal N}=1,2$ supersymmetric gauge theories, Eq.\,(\ref{cc})
 shows that in general $a\neq c$. Thus the terms relevant to
 gravitational background in the superconformal  anomaly
  should arise from the supergravity side. In Ref.\,\cite{ahar},
  it was shown that for the ${\cal N}=2$ supersymmetric $USp(2N)$
  gauge theory coupled to two hypermultiplets in the fundamental and
  antisymmetric tensor representations of the gauge group, respectively,
  the gravitational background part in the holographic chiral anomaly
  does originate from a mixed CS term. However, this CS
  terms is obtained from the compactification on $S^3$ of the Wess-Zumino
  term describing the interaction of the R-R 4-form field with
  eight $D7$-branes and one orientifold 7-plane system. Specifically,
  this gravitational background term is at the subleading $N$ order
  rather than the leading $N^2$ order in the large-$N$ expansion
  of  ${\cal N}=2$ supersymmetric $USp(2N)$ gauge theory.
  Due to the supersymmetry, there should also arise the $N$-order
  part containing the $R_{\mu\nu\lambda\rho} R^{\mu\nu\lambda\rho}$
  term in the holographic Weyl anomaly. This fact was explicitly shown
  in Ref.\,\cite{blau}. The authors first performed $T$-duality
  and realized that the above mixed CS term  originates from
  Green-Schwarz couplings in the heterotic or type I supergravity.
  Further,  they  considered the higher-order gravitational
  terms  required by the supersymmetry
  in the low-energy effective action of one-loop
  heterotic or type $I$ string theory \cite{berg2} and then performed
  the heterotic-type $I$-type $I'$ T-duality to extract the
  desired  gravitational term in the $AdS_5$ supergravity. Finally,
  the holographic Weyl anomaly containing the
  square of the Riemannian tensor
  was found  along the line of
  Ref.\,\cite{hesk}. These two cases imply that
  we should consider the full supersymmetric action
  consisting of the higher-order gravitational terms and
  its supersymmetric variation in the $AdS_5$ background will
  most probably lead to
  the term (\ref{gc}) in the holographic super-Weyl anomaly. However,
  the brane configuration for the ${\cal N}=1$ $U(N)$ gauge
  theory is quite complicated, being a $D4-D6-NS5-NS5'$ system
  \cite{herb}. It is not straightforward to write
  down the mixed CS term from the
  corresponding Wess-Zumino term
   and to extract the explicit form of the supersymmetric action
  containing the higher-order gravitational terms. We hope
  to make a complete investigation along the line similar to the one
 adopted in Refs.\,\cite{ahar,blau}.

The other possible reason for the failure of getting the
 gravitational background
contribution is that in Eqs.(\ref{rgf1}), (\ref{redg}),
(\ref{redsl}) and (\ref{redsr})
 we  consider only the leading order of $r$-dependence
of the bulk fields near the $AdS_5$ boundary. This argument is
inspired by the process of deriving the holographic Weyl anomaly.
It was shown in Ref.\,\cite{hesk} that
 if one makes a complete near-boundary
analysis and considers the asymptotic expansion beyond the leading
order up to logarithmic term \cite{hesk,bian1}, i.e.,
\begin{eqnarray}
{\cal F}(x,r)=r^m\left[ f_{(0)}(x)+ f_{(2)}(x) r+\cdots r^n \left(
f_{(2n)}(x)+ \widetilde{f}_{(2n)} (x)\,\ln r +\cdots\right)+\cdots
\right],
\end{eqnarray}
the higher-order gravitational terms can arise in the on-shell
action
 \cite{hesk}, and they lead to the holographic Weyl anomaly composed
of the $R_{\mu\nu} R^{\mu\nu}$ and $R^2$ terms. Therefore, it is
also possible that  the gravitational background field part in the
super-Weyl anomaly can arise as the holographic Weyl anomaly does
\cite{hesk} if one takes into account the logarithmic terms in
above expansion. Of course, in this case the on-shell gauged
five-dimensional supergravity action has the infrared divergence
near the $AdS_5$ boundary due to its infinite boundary. One must
perform a holographic renormalization procedure to make it well
defined \cite{bian1}.  However, although we do not exclude that
this is a possible source, it seems that the first reasoning is a
more convincing one.

%Actually,  This can also give only the contribution from the external
%gauge field.

 %To achieve the contribution to the chiral R-symmetry anomaly
 %from the gravitational
 %background, one must consider the asymptotic expansion of the bulk fields
 %until the logarithmic function  of the radial coordinate
 %emerges. The superficial  reason  for this  lies in the
 %existence of the Chern-Simons term
 %consisting only of the vector field.
 %The Chern-Simons term is a topological term and becomes
 %a total derivative under gauge transformation. Its supersymmetric
 %variation is not a total derivative term, but to achieve supersymmetry
 %for the bulk gauged supergravity, as shown in
 %Eqs.\,(\ref{var1})--(\ref{var3}), one must choose the indefinite
 %coefficient of the Chern-Simons term to make its non-derivative
 %supersymmetric variation  cancel with the other one. In particular,
 %unlike the induced metric $l^2 g_{\mu\nu}(x,r)/r^2$ and
%the gravitino $\Psi_\mu (x,r)$,
 %the leading term in the near-boundary asymptotic expansion
 %of the gauge field ${\cal A}_\mu (x,r)$ is
 %independent of the radial coordinate.
 %In Eq.\,(\ref{var3}), it is the
 %total derivative term relevant to the supersymmetric variation
 %of the Chern-Simons term that leads to the external gauge field
 %contribution to the super-Weyl anomaly of supersymmetry current.
 We are not aware  whether there exist any
 physical reasons for the difference between these two holographic
  contributions
  to the superconformal anomaly. Since the essence of the holographic
  anomaly
  is the anomaly inflow from the bulk theory \cite{callan},\footnote{
  We thank  A. Kobakhidze for discussion on this point.}
  this is the reason why one can extract out the anomaly  from a
  higher-dimensional bulk theory,
  Thus it might be relevant to the difference between the
  anomaly inflows contributed by the gravitational and gauge field
  backgrounds.

Finally, it is worth to emphasize another way of calculating the
holographic Weyl anomaly proposed in Ref.\,\cite{imbi}. This
approach is independent of the concrete form of the classical bulk
gravitational action, and depends  purely on a special bulk
diffeomorphism (called the `` PBH '' transformation \cite{pbh}),
which keeps the form of the Fefferman-Graham metric \cite{feff} of
an arbitrary $d+1$-dimensional manifold with the boundary
topologically isomorphic to $S^d$ invariant and reduces to a Weyl
transformation on the boundary. With a choice on the holomorphic
dimensional regularization, there emerges no logarithmic  function
of the radial coordinate in the near-boundary asymptotic
expansion. Further, the invariance of a
 general bulk gravitational action, which admits an $AdS_{d+1}$
 classical solution,
 under this particular diffeomorphism  yields a
Wess-Zumino-like consistency condition satisfied by the generating
functional for the Weyl anomaly on the boundary ($d$ being an even
integer).  Hence the hologrpahic Weyl anomaly can be extracted out
and no holographic renormalization procedure is necessary
\cite{imbi}. The advantage of this approach is that one can avoid
the complicated near-boundary analysis and the subsequent
holographic renormalization procedure. One may consider to use
this approach to calculate the super-Weyl anomaly of a
supersymmetry current. However, there is one crucial obstacle to
overcome. That is, one needs to find a supersymmetric
generalization of above special bulk diffeomorphism, which should
keep both the supersymmetry in the bulk and on the boundary. In
such an approach, one must employ the holographic dimensional
regularization to prevent the logarithmic dependence from emerging
in the near-boundary asymptotic expansion. However, the
dimensional regularization in general does not preserve the
supersymmetry. This fact has cast a shadow on the application of
the above approach to the evaluation of the holographic
supercurrent anomaly.

\bigskip

\acknowledgments \noindent We are indebted to M. Grisaru,
 A. Kobakhidze, M. Nishimura and A. Schwimmer
for discussions. We also would like to thank G. Kunstatter, R.G.
Leigh,
 C. Montonen and M. Sheikh-Jabbari for useful remarks.
 This work is supported by
the Academy of Finland under the Project No. 163394.

\end{document}